\documentclass[aps,prb,reprint,superscriptaddress,amssymb,showpacs,showkeys,floatfix]{revtex4-1}

\usepackage{graphicx,amsmath,bm,xcolor}

\newcommand{\uu}[1]{\ensuremath\,\mathrm{#1}}

\begin{document}

\date{\today}
\title{First principles study of water-based self-assembled nanobearing effect in CrN/TiN multilayer coatings}

\author{David Holec}
\email{david.holec@unileoben.ac.at}
\affiliation{Department of Physical Metallurgy and Materials Testing, Montanuniversit\"at Leoben, Franz-Josef-Strasse 18, A-8700 Leoben, Austria}
\author{J\"org Paulitsch}
\email{joerg.paulitsch@unileoben.ac.at}
\affiliation{Department of Physical Metallurgy and Materials Testing, Montanuniversit\"at Leoben, Franz-Josef-Strasse 18, A-8700 Leoben, Austria}
\author{Paul H.~Mayrhofer}
\affiliation{Department of Physical Metallurgy and Materials Testing, Montanuniversit\"at Leoben, Franz-Josef-Strasse 18, A-8700 Leoben, Austria}

\begin{abstract}
Recently, we have reported on low friction CrN/TiN coatings deposited using a hybrid sputtering technique. These multilayers exhibit friction coefficients $\mu$ below 0.1  when tested in atmosphere with a relative humidity $\approx25\%$, but $\mu$ ranges between $0.6$--$0.8$ upon decreasing the humidity below $5\%$. Here we use first principle calculations to study O and H adatom energetics on TiN and CrN (001) surfaces. The diffusional barrier of H on TiN(001) is about half of the value on CrN(001) surface, while both elements are stronger bonded on CrN. Based on these results we propose a mechanism for a water-based self-assembled nanobearing.
\end{abstract}



\maketitle


Chromium and titanium nitride (CrN, TiN) thin films are widely used as hard protective coatings for various industrial in automotive applications as they show high hardness and increased wear and corrosion resistance \cite{SUNDGREN1985,MUSIL1988,Musil1993,Jensen1993,Hurkmans1996,Hurkmans1999}. Tribological investigations of such coatings indicate that the coefficient of friction $\mu$, is around 0.45 and 0.8 for CrN and TiN coatings, respectively \cite{Lee2004,Badisch2003,Stoiber2003,Kacsich1996,Takadoum1997}. \citeauthor{Ehiasarian2004} and \citeauthor{Paulitsch2007} showed that using high ionizing deposition techniques like the high power impulse magnetron sputtering (HIPIMS) for depositing these coatings, leads to increased wear resistance and reduced $\mu$ values due to the formation of dense coating structures \cite{Ehiasarian2004,Paulitsch2007,PAULITSCH2008,Paulitsch2010a}. Nevertheless, $\mu$ values below 0.1, comparable to diamond-like carbon coatings or carbo-nitrides, could not be achieved.

\begin{figure}[b]
  \includegraphics[width=\columnwidth]{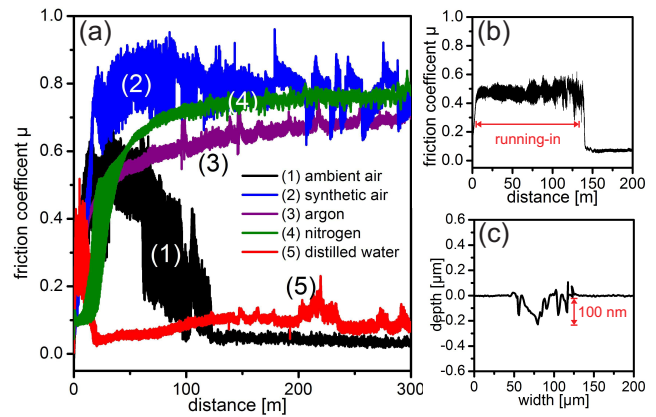}
  \caption{(colour online): (a) BOD tests of CrN/TiN superlattice coatings in different atmospheres, published in Ref. \onlinecite{Paulitsch2010}. (b) Evaluation of the running-in length, measured in ambient air with a relative humidity of around 25\%, and (c) the resulting wear track depth of a CrN$_{\mathrm{MPP}}$/TiN$_{\mathrm{DCMS}}$ multilayer coating with a bilayer period of $10\uu{nm}$, after stopping the BOD test in the low friction steady state regime.}\label{fig1}
\end{figure}

Recently, we have deposited multilayer coatings of CrN and TiN by simultaneously sputter the metal Cr target in HIPIMS or modulated pulse power (MPP) mode, and the metal Ti target in direct current magnetron sputtering (DCMS) mode \cite{Paulitsch2012,Paulitsch2010}. The resulting films indicate a dense superlattice structure with a bilayer period $\lambda$ from 6 to $10\uu{nm}$, hardness values $\approx25\uu{GPa}$, and a preferred (001) orientation \cite{Paulitsch2012,Paulitsch2010}. Tribological investigation using a ball-on-disk (BOD) tribometer yielded wear rates $\approx 3\cdot10^{-16}\uu{m^3/Nm}$ and a coefficient of friction below 0.1 when tested at room temperature (RT) and relative ambient humidity of around 25\% \cite{Paulitsch2012,Paulitsch2010} (see Fig.~\ref{fig1}a, curve (1)). Investigations of the triggering effect for the low friction values, by evaluating the wear depth after stopping the BOD testing when the $\mu$ value drops below 0.1, showed that a polishing-in depth of around $100\uu{nm}$ is necessary, see Fig.~\ref{fig1}b and c. Furthermore, variations of the ambient air during testing by introducing dry argon, nitrogen or synthetic air, which all reduce the relative humidity to values below 5\%, as well as tests in a water bath indicate that the low friction effect of the CrN/TiN multilayer coatings depends sensitively on the relative humidity during testing (see Fig.~\ref{fig1}a curves (2) to (5))\cite{Paulitsch2010}.

\begin{figure*}[t!]
  \includegraphics[width=0.74\textwidth]{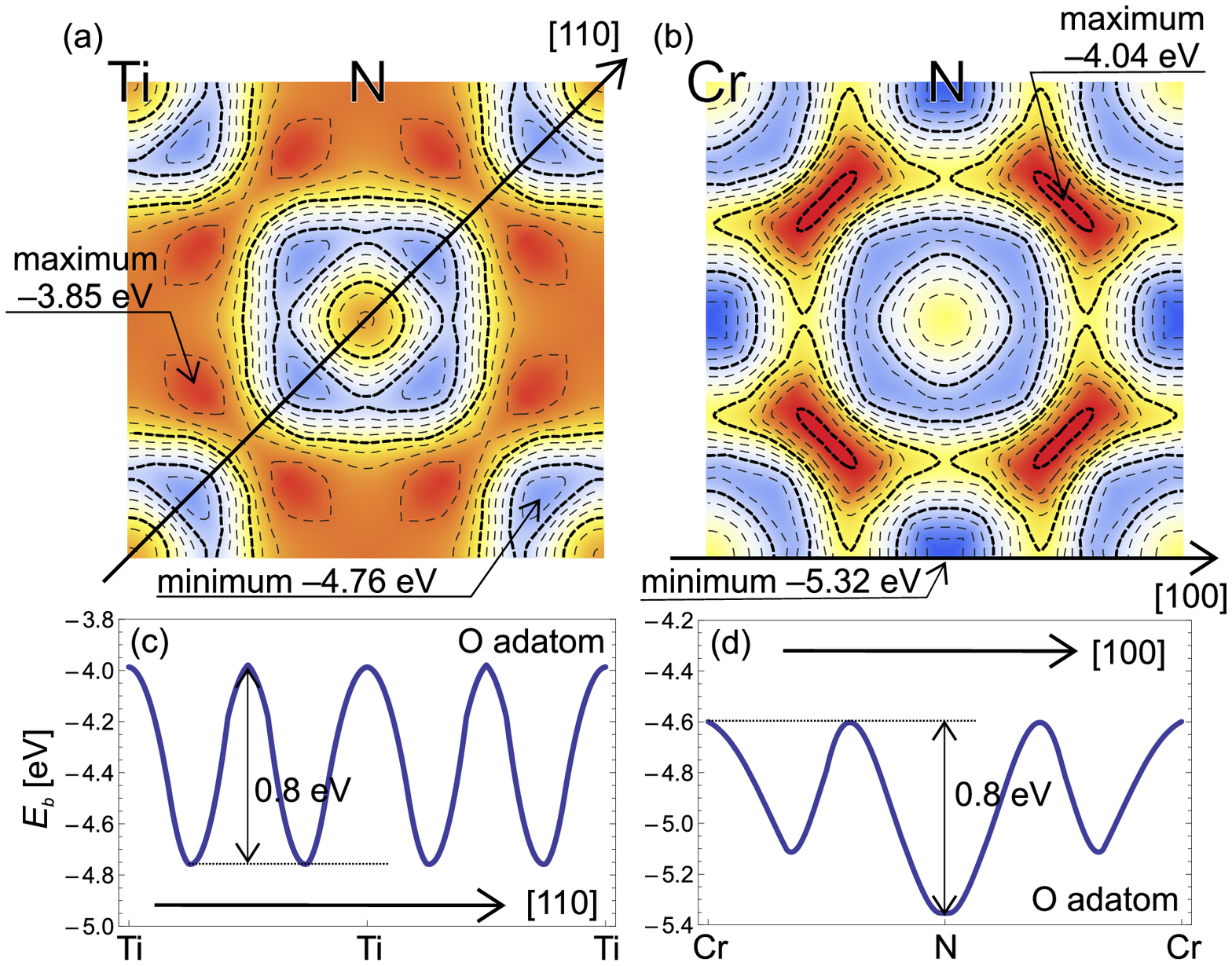}
  \caption{(colour online): Potential energy surface of O adatom on (001) surface of (a)~TiN and (b)~CrN. The 2D cuts in (c)~TiN$\langle110\rangle$ and (d)~CrN$\langle100\rangle$ directions show the maximum energy barrier for surface diffusion.}\label{fig2}
\end{figure*}

The aim of this study is to clarify the above mentioned observations of water rather then oxygen being the essential ingredient to obtain the low friction coefficient. In this Letter we report on density functional theory (DFT) calculation of the hydrogen and oxygen adatom interactions with the free CrN and TiN surfaces, as a first approach to the complex interaction between the water molecule and multilayer system. We employed Vienna ab initio Simulation Package (VASP) \cite{Kresse1996a,Kresse1993} together with projector augmented wave (PAW) pseudopotentials \cite{Kresse1999} using the generalised gradient approximation (GGA) as parametrised by \citet{Wang1991}. The reciprocal space was sampled with minimum of $8000\,\bm{k}\mbox{-points}\cdot\mbox{atom}$ and the plane wave cutoff energy was $450\uu{eV}$. The antiferromagnetic configuration of cubic CrN (B1, NaCl prototype) was modelled as layers of alternating spins (afm0). Although the true ground state, afm1, has a slightly different arrangement \cite{Corliss1960,Miao2005,Mayrhofer2008,Alling2010a}, the energy of formation, lattice parameters and bulk modulus of these two configurations are very similar (e.g. $\Delta E_f^{\mathrm{afm0}-\mathrm{afm1}}=6\uu{meV/atom}$ or $\Delta B_0^{\mathrm{afm0}-\mathrm{afm1}}=2\uu{GPa}$). Thus we used the afm0 configuration in all our calculations since it is considerably less computationally demanding than the afm1 due to a smaller unit cell.

In order to calculate the potential energy surface (PES) for the adatom diffusion on the (001) surfaces, we first optimised the slab and vacuum thicknesses ($\approx25\uu{\mbox{\AA}}$ and $12\uu{\mbox{\AA}}$, respectively) for getting converged surface energies. The procedure yielded $60\uu{meV/\mbox{\AA}^2}$ for CrN(001) and $81\uu{meV/\mbox{\AA}^2}$ for TiN(001), the latter value corresponding to those reported in literature \cite{Siodmiak2001,Gall2003}. Subsequently, we used the same slab geometry with an adatom, and for a dense grid of point spanning the (001) surface we optimised the total energy of the system by adjusting the adatom’s distance from the surface (with fixed lateral coordinates). The binding energy, $E_b$ of an adatom reads:
\begin{equation}
  E_b=-(E_{\mathrm{total}}^{\mathrm{slab}+\mathrm{adatom}}-E_{\mathrm{total}}^{\mathrm{slab}}-E_{\mathrm{total}}^{\mathrm{adatom}})\ .
\end{equation}

Figure~\ref{fig2} shows the PES of oxygen adatom on (001) surface of TiN and CrN. An inspection of the absolute values reveals that O is stronger bonded to the CrN surface ($E_{b,\max}\approx5.3\uu{eV}$) than on the TiN surface ($E_{b,\max}\approx4.8\uu{eV}$). Oxygen atoms are strongly bonded in the vicinity of the Ti and Cr sites. The O adatoms are strongly bonded also to the N sites on the TiN(001) surface, while the binding is very weak above N sites on CrN(001) surface (cf. Figs.~\ref{fig2}a and b). The lowest energy barrier (from the PES minimum) for the surface diffusion of O on the TiN surface is $\approx0.8\uu{eV}$, corresponding to a movement along the $\langle110\rangle$ directions, thus suggesting a zig-zag movement between Ti sites and avoiding N sites. The lowest diffusion barrier for O on CrN is also approximately $0.8\uu{eV}$, however here in the $\langle100\rangle$ directions. Consequently, oxygen atoms come during the diffusion to the vicinity of both, Cr and N atoms. The diffusional behaviour of O adatoms is therefore qualitatively different on CrN and TiN (001) surfaces.

\begin{figure*}[t!]
  \includegraphics[width=0.74\textwidth]{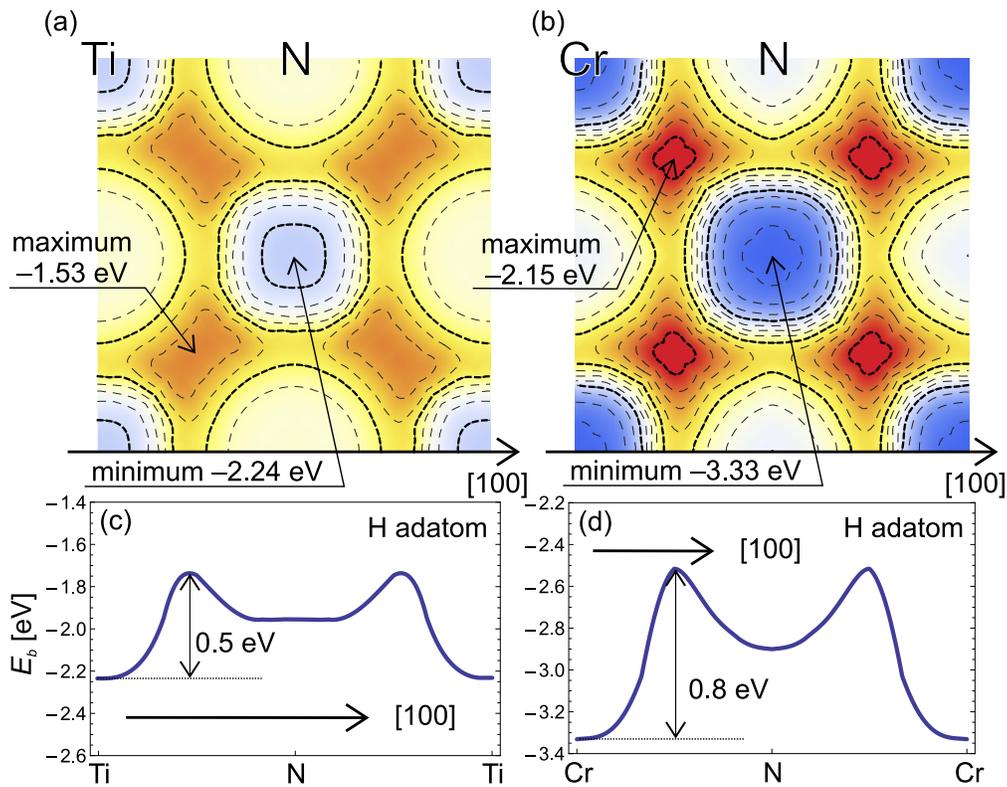}
  \caption{(colour online): Potential energy surface of H adatom on (001) surface of (a)~TiN and (b)~CrN. The 2D cuts in (c)~TiN$\langle100\rangle$ and (d)~CrN$\langle100\rangle$ directions show the maximum energy barrier for surface diffusion.}\label{fig3}
\end{figure*}

The energetics of H adatom on TiN and CrN (001) surfaces is shown in Fig.~\ref{fig3}. In contrast to the O behaviour, hydrogen PES is qualitatively the same for both materials. In both cases, the energetically preferred adatom site is above the Ti or Cr atoms, whereas N sites exhibit local minima in PES. The lowest energy barriers for diffusion are along the $\langle100\rangle$ directions, suggesting that H atoms come close to both, Ti or Cr and N sites during surface diffusion. Inspection of the PES profiles along the $\langle100\rangle$ direction, however, reveals that the diffusion barrier is $\approx0.5$ and $\approx0.8\uu{eV}$ on TiN and CrN surfaces, respectively. As a consequence, H atoms are predicted to be more mobile on TiN(001) than on the CrN(001) surface.

The previous findings may be summarised as follows: (i)~H and O are stronger bonded on the CrN than on TiN (001) surface, (ii)~the diffusion barriers for O are comparable on both materials, and (iii)~H diffuses much easier on the CrN surface than on TiN (the diffusion barrier on the CrN surface is about half of that on TiN). Based on these results one can speculate about the behaviour of a water molecule, as an entity bonded either via O atom or via H atom to the CrN/TiN surface: the water molecule is expected to be more mobile on TiN surface (due to the smaller diffusion barrier for H atoms) and to be stronger bonded on the CrN. As a consequence of the multilayer (bi-material) arrangement of the CrN/TiN coatings, the water may spontaneously concentrate on the CrN layers while it depletes on the TiN layers, thus acting as a self-assembled nanobearing. Such mechanism is indirectly supported also by the fact, that in order to get into the low-friction mode, a defined running-in distance is first needed (see Fig.~\ref{fig1}). This corresponds to the development of a wear track spanning over several layers (e.g., wear track depth of $\approx100\uu{nm}$ for the bi-layer period $\lambda=10\uu{nm}$); only after a certain number and geometry of layers is exposed to the counterpart surface, the self-assembly of water droplets takes place to promote the low friction.

In conclusion, we have reported on diffusional properties of H and O adatoms on TiN and CrN(001) surfaces. H is shown to be more mobile on TiN, O exhibits the same diffusion barriers on both surfaces. Both elements are stronger bonded on the CrN than on the TiN surface. Subsequently, we used these results to speculate about the behaviour of water molecules on the CrN/TiN multilayer surface that would rationalise our experimental observations. We propose that the water droplets in the wear track exhibit a tendency for self-assembly with a nanobearing-like effect.

The authors are grateful to the financial support by the START Program (Y371) of the Austrian Science Fund (FWF).

\bibliographystyle{aipnum4-1.bst}
\bibliography{surface_diffusion_crn-tin.bib}
\end{document}